**Bottom-up strategies for the assembling of magnetic systems using nanoclusters.**

*Authors: V. Dupuis\*, A. Hillion, A. Robert, O. Loiselet, G. Khadra, P. Capiod, C. Albin, O. Boisron, D. Le Roy, L. Bardotti, F. Tournus, A. Tamion*

Institut Lumière Matière, UMR 5306, Université Lyon 1-CNRS, Université de Lyon, 69622 Villeurbanne cedex, France

\*: email address of the corresponding author: veronique.dupuis@univ-lyon1.fr

**Abstract**

In the frame of the 20th Anniversary of the Journal of Nanoparticle Research (JNR), our aim is to start from the historical context twenty, years ago and to give some recent results and perspectives concerning nanomagnets prepared from clusters preformed in the gas phase using the Low Energy Cluster Beam Deposition (LECBD) technique. In this paper, we focus our attention on the typical case of Co clusters embedded in various matrices to study interface magnetic anisotropy and magnetic interactions as a function of volume concentrations, and on still current and perspectives through two examples of binary metallic 3d-5d TM (namely CoPt and FeAu) clusters assemblies to illustrate size-related and nanoalloy phenomena on magnetic properties in well-defined mass-selected clusters. The structural and magnetic properties of these cluster assemblies were investigated using various experimental techniques that include High Resolution Transmission Electron Microscopy (HRTEM), Superconducting Quantum Interference Device (SQUID) magnetometry, as well as synchrotron techniques such as Extended X-Ray Absorption Fine Structure (EXAFS) and X-Ray Magnetic Circular Dichroism (XMCD). Depending on the chemical nature of both NPs and matrix, we observe different magnetic responses compared to their bulk counterparts. In particular, we show how finite size effects (size reduction) enhance their magnetic moment and how specific relaxation in nanoalloys can impact their magnetic anisotropy.

**Introduction**

In the last two decades, magnetic nanoparticles have attracted much attention both for their fundamental interest and for their potential applications. Nowadays, it is well known that at such nanosize, the magnetization direction of a particle usually fluctuates at room temperature and that such *superparamagnetic* behaviour is a crucial issue in fundamental research as well as practical applications (Mornet et al. 2004, Carrey et al. 2006, Binns 2013). But in the end of the 90's, with the beginnings of the synthesis of cluster-assembled nanostructures by Low Energy Cluster Beam Deposition (LECBD) methods (Perez et al. 1997, Milani and Iannota 1999), it was the starting point for the magnetic properties study of supported transition metal (TM) clusters.

Since 1985, first results have been reported on Stern-Gerlach magnet deflection from spatially resolved time-of-flight photoionization mass spectrometry measurement of small free-clusters beam (Cox et al. 1985) where isolated iron clusters ranging in size from 2 to 17 atoms, have been observed like-paramagnetic. Five years later, again from Stern-Gerlach deflections of a molecular beam in an improved laser vaporization source (Milani 1990), (De Heer et al. 1990) measured magnetic moments for cold iron clusters, with number of atoms per cluster $N$ up to 650, but claimed that the average clusters magnetic moments are in all cases below the bulk value and increase with a "cluster internal temperature". The next year, their results conflicted with the first experimental evidence by (Bucher et al 1991) of the *superparamagnetic* model for isolated monodomain nanoparticles predicted by (Linderoth and Khanna 1992), where the concept of rapid orientation fluctuations of the "true magnetic moment" was introduced. Then a series of intense experimental works on clusters in flight helped in showing that the magnetic moment of 3d-TM clusters always exceeds the bulk value up to *N*= 500 atoms and progressively decreased with *N* and temperature (Billas et al. 1993). Meanwhile the appearance of giant magnetic moment in 4d-TM clusters as Rhodium has been found for *N*<40 atoms by (Cox et al. 1993). In that regard, calculations based nowadays on an electronic spin-fluctuation theory and a parallel tempering Monte Carlo approach, have confirmed the observed strong stability of the ferromagnetic (FM) order within small clusters and described finite-temperature magnetic properties of Fe-clusters as a function of size, structure, and interatomic distances (Garibay-Alonso et al. 2009).

Nevertheless, to experimentally observe such finite-size dependent magnetic effects on supported clusters, it has been necessary to develop alternative techniques to avoid magnetic interactions between deposited clusters on substrate and contamination with external environment in particular upon transfer in air. Then, some pioneer groups started to prepare assemblies of magnetic nanostructure by using the original bottom up LECBD technique which ensures the direct soft-landing under high vacuum conditions (HV) of very low energy clusters produced in an inert-gas condensation source. In the regime of low kinetic energy deposition, incident free clusters (kinetic energy $E_c$ < 1 eV/atom) should not fragment upon impact on the substrate (Haberland et al. 1995) leading to the formation of cluster-assembled dots or films which retain partial original structures and properties of gas-phase clusters (Paillard et al. 1993, Jensen 1999). Moreover, a co-deposition of cluster and atomic independent beams simultaneously on a same substrate, allow us to produce any kind of cluster/matrix system, even with miscible elements, in a wide range of cluster volume concentration (from 0.1% up to 60%) (Bansmann et al. 2005). As pioneer results, one can mention *superparamagnetic* like-behavior of cobalt clusters assemblies embedded in silver and $SiO_x$ matrix with independent control of the cluster size and concentration (Dupuis et al. 1997). In this case, the concentration dependence of giant magnetoresistance (GMR) of cobalt clusters embedded in silver matrix has been studied in such granular films (F. Parent et al. 1997). Optimal GMR values of more than 10 % has been achieved but only in highly concentrated samples close to percolation threshold, where the simple quadratic relation ($\Delta R$ α $(M/M_{sat})^2$) between magnetic and magnetoresistive response is indeed no longer observed and the model of (Allia et al. 1995) is no longer valid (Oyarzun et al. 2013).

In parallel, progress in spectroscopy experiments under synchrotron radiation (SR) facilities of 3rd generation (Mülhaup 1995), such as X-ray magnetic circular dichroism (XMCD) started to provide a more detailed description of the magnetic properties of clusters deposited on surface or embedded in a matrix. The crucial advantage of XMCD is to be able to separate the spin *($\mu_s$)* and orbital *($\mu_L$)* contributions of the total cluster magnetic moment *($\mu_N$)* with the chemical selectivity of the SR. Thus, the quantitative $\mu_s$ enhancement by reducing *N*, has been confirmed in supported 3d TM-clusters (see references in Bansmann et al. 2005, Ohresser et al. 2013, Binns 2013). Furthermore, the direct access to the average $\mu_L$ started to give a clear insight of the role of spin-orbit interactions in small supported clusters (Edmonds et al. 1999) on the origin of magnetic anisotropy energy (MAE). Note that the MAE defined as the energy involved in the spontaneous magnetic moment reversal from an easy axis to the opposite one, determine the *Blocking* temperature above which *superparamagnetic* behavior sets. Thus energy barrier has been accurately quantitatively determined later from high sensitivity magnetometry measurements on single nanomagnet or well-separated clusters assemblies (Jamet et al. 2001, Tamion et al. 2009, Pierron-Bohnes et al. 2012).

Nowadays, a huge progress in synthesis such as mass-selected LECBD synthesis techniques under Ultra-High Vacuum (UHV) (Tournus et al. 2011) and in characterization as aberration-corrected transmission electron microscopy in combination with spectroscopic methods, enables atomic resolution structural characterization (Pohl et al. 2014). In this paper, we give an overview of experimental results obtained on Co clusters embedded in various matrices to study interface magnetic anisotropy and interactions, and open questions and perspectives through two examples of binary metallic 3d-5d (namely CoPt and FeAu) clusters assemblies to illustrate size-related and nanoalloy phenomena on magnetic properties in well-defined mass-selected clusters.

## I- Pure Co Clusters embedded in various matrices and concentrations

Clusters are produced in a laser vaporization–gas condensation source similar to that developed by Smalley and improved by (Milani and de Heer 1990). Briefly, a plasma created by the impact of a frequency doubled Nd:YAG (Yttrium Aluminium Garnet) laser beam focused on a rod, is thermalized by injection of a continuous flow of helium at low pressure (typically 30 mbar) inducing cluster growth. Clusters are subsequently stabilized and cooled down in the supersonic expansion taking place at the exit nozzle of the source. A low energy cluster beam is then obtained, with clusters in the nanosize range (1.5-8 nm) which are codeposited under ultrahigh vacuum (UHV) conditions ($10^{-10}$ mbar in base pressure condition) simultaneously with the independent atomic beam of the matrix.

When decreasing the size of magnetic particles down to the nanometric range, the Magnetic Anisotropy Energy (MAE) which acts to fix the magnetization along an easy axis, is counterbalanced by the thermal energy $k_B T$ resulting in magnetization fluctuations between opposite directions. Generally, diluted assemblies of magnetic nanoclusters are expected to present two different magnetic states, as a function of temperature. The transition from the ferromagnetic blocked regime to the *superparamagnetic* one, above the so-called blocking temperature ($T_B$), can be evidenced as a lack of hysteresis in the magnetic loops and a peak in the Zero Field Cooled (ZFC) curves. $T_B$ is defined as the temperature where the relaxation time: $\tau(T) = \tau_0 \exp(\Delta E / k_B T)$ for magnetization reversal (with $k_B T$ the thermal energy and $\tau_0^{-1}$ the attempt frequency - typically in the range of $10^9$-$10^{12}$ Hz) is comparable to the measuring time $\tau_{mes}$. In this case, $\Delta E$ corresponds to the energy barrier to overcome in order to reverse the particle magnetization i.e. to the MAE in zero magnetic field. Nevertheless, whereas the magnetic moment is proportional to the cluster magnetic volume, the cluster size dependence of $\Delta E$ is more complex. Because our nanoparticles present a high surface-to-volume ratio, a careful analysis is essential to distinguish the interfacial and the volume contributions.

### *Assemblies of Co-clusters in C, Cu and Au matrices*

Very recently, we have reported on the magnetic properties of Co clusters embedded in different matrices (C, Cu and Au) (Hillion et al. 2017). The samples are formed by Co nanoparticles around 2.5 nm in diameter determined by transmission electron microscopy (TEM), and their diameter probability density function (PDF) closely follows a lognormal distribution. In order to measure the clusters magnetic intrinsic properties from Superconducting Quantum Interference Device (SQUID) magnetometry, we have prepared highly diluted samples. By using Isothermal Remanent Magnetization (IRM) and Direct current Demagnetization (DcD) curves at 2 K (described in Hillion et al. 2013) and from the following equation: *Δm = DcD(H)-(IRM(∞)-2IRM(H))*, we have verified that magnetic interactions are negligible as the *Δm* parameter is found equal to 0 whatever the applied magnetic field for volume concentration lower than 1%.

Then we extended the already powerful "triple" fit approach, where the Zero field Cooled/Field Cooled (ZFC/FC) susceptibility curves and a *superparamagnetic* magnetization loop are simultaneously fitted with a semi-analytical model (Tamion et al. 2009), to the low-temperature hysteresis loop and IRM curve which bear distinct signatures of the particles magnetic properties (Tamion et al. 2012). Briefly, because at the cluster surface, the cubic symmetry is broken, the anisotropy function of a macro-spin involves second-order dominating terms, and can be expressed as:

$$G(\theta, \varphi) = K_1 m_z^2 + K_2 m_y^2,$$

with z the easy axis, y the hard axis and $K_1<0<K_2$. Here $K_1$ and $K_2$ represent the second order anisotropy constants, $m_z$ the normalized magnetization projection on the easy axis. Finally $\vartheta$ and $\varphi$ represent the magnetization angles in a spherical basis. In a case of a biaxial anisotropy we use the geometric approach to build the astroid which represents, in the field space, the magnetic switching field ($H_{sw}$). To take into account the thermal fluctuations which can bring the magnetization over the energy barrier; we use the Néel's relaxation model (Néel 1949). When two stable positions exist the relaxation time between these states is given by the previous defined equation: $\tau(T) = \tau_0 \exp(\Delta E/k_B T)$. It is therefore possible to simulate hysteresis loops of an assembly of nanoparticles taking into account the temperature, the size distribution and clusters' biaxial anisotropy (see Fig. 1).

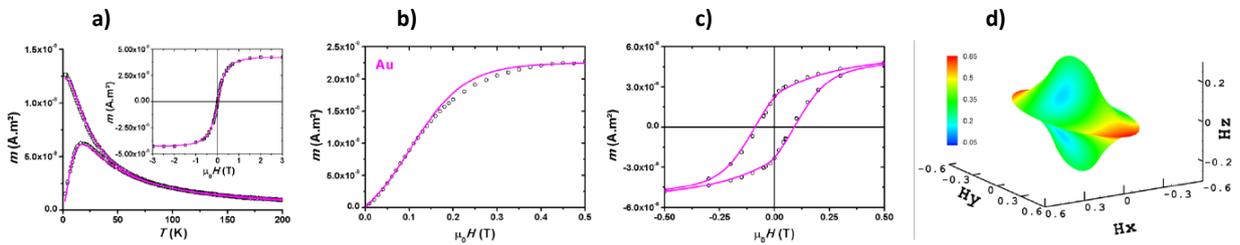

**Fig. 1:** ZFC/FC susceptibility curves taken at 5 mT and magnetization curves at 300 K in insert (a), IRM curves (b) and Hysteresis loops (c) at 2 K for Co nanoparticles embedded in Au matrix (0.5 % vol.). The solid lines correspond to the adjustments using the same set of parameters. The corresponding 3D astroid associated to the biaxial fit is shown at the right (d).

For each Co sample with C, Cu and Au matrix, we have reached a reliable determination of the ZFC peak temperature ($T_{max}$), the low-temperature coercivity ($\mu_o H_c$), the median magnetic diameter ($D_m$) and its dispersion parameter ($\omega$) as a log-normal NPs size distribution and the MAE normal distribution (characterized by the median values $K_1$, $K_2$ and the standard deviation $\sigma_{K1}$), all reported in Table I. First of all, we can notice that the smallest magnetic diameter ($D_m$) is obtained for carbon matrix while the largest ($D_m$) is found for Co@Au samples. In one hand in as-prepared Co@C samples, a magnetically dead shell layer has been already attributed to a metastable carbide formation at the interface between Co-fcc core cluster and carbon matrix (leading to a $\mu_s$ decrease at Co environment from XMCD measurements) but which can be removed by HV annealing at 750 K without deteriorating the PDF or changing the magnetic anisotropy constant (Tamion et al. 2011).

|  | Co@C | Co@Cu | Co@Au |
|---|---|---|---|
| $D_m$ (nm) | 2.1 ± 0.2 | 2.5 ± 0.2 | 2.8 ± 0.2 |
| $T_{max}$ (K) | 6.5 | 12 | 17 |
| $\mu_o H_c$ (mT) | 25 | 40 | 85 |
| $\omega$ | 31% ± 5% | 27% ± 5% | 26% ± 5% |
| $K_1$ (kJ.m$^{-3}$) | 115 ± 10 | 155 ± 10 | 190 ± 10 |
| $\sigma_{K1}$ (kJ.m$^{-3}$) | 40 ± 10 | 62 ± 7 | 90 ± 10 |
| $K_2/K_1$ | 1.2 ± 0.1 | 1.2 ± 0.1 | 1.3 ± 0.1 |

**Table I**: Magnetic characteristics of Co clusters diluted in C, Cu and Au matrix with 2.5 nm in diameter from (Hillion et al. 2017).

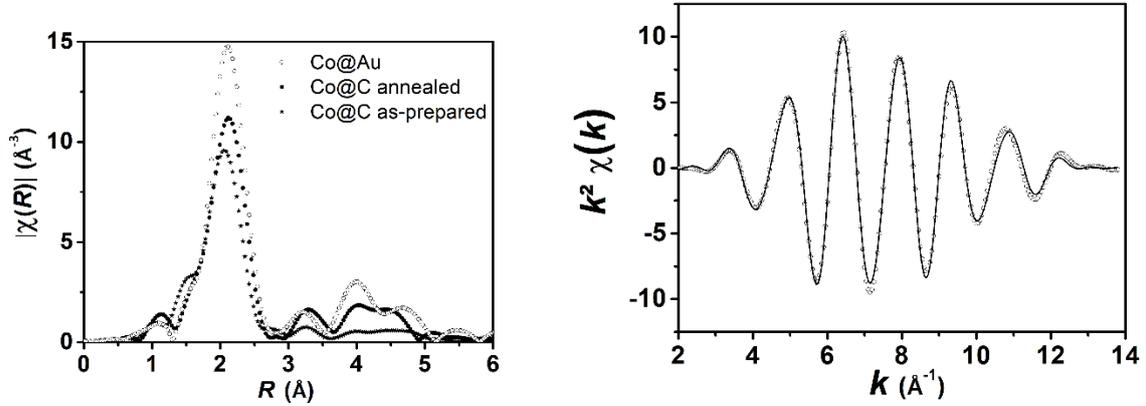

**Fig. 2:** Fourier Transform of the EXAFS signal for Co clusters embedded in C (as-prepared and annealed) and in Au matrix with 2.5 nm in diameter (left). Comparison between the experimental EXAFS signal (dots, contribution of the nearest neighbours (NN) peak only) and simulated curves (solid lines) at the Co-K edge for Co@Au sample.

This interpretation is confirmed by experiments at the Co-K edge (7713 eV) performed at room temperature on the CRG-BM30b-FAME beam line of the ESRF (Proux et al. 2006) where a significant increase in the amplitude of the Fourier Transform of the extended x-ray absorption edge fine structure (EXAFS) oscillations is observed for Co@C sample upon annealing (see figure 2, left). On the other hand, the FT EXAFS oscillations is very well defined up to 6Å with a metallic bulk like-profile for Co@Au sample. For this sample, from a quantitative EXAFS analysis using the Artemis software, the best fit for the nearest neighbours (NN) peak has been obtained by considering only two first-shell environments, namely Co-Co and Co-Au with coordination numbers NN equal to 8.5 and 1.9 and average interatomic distances R equal to 2.48 Å and 2.66 Å, respectively. As comparison, R is equal to 2.507 Å in Co-bulk and to 2.68 Å for CoAu speculative alloys from a Vegard law. Due to finite size effects, the total coordination number NN equal to 10.4 is reduced (compared to 12 in Co-bulk) because 45 % of Co atoms for 2.5 nm Co-cluster are located in the first surface-monolayer. Nevertheless, the NN ratio of 17% between both environments Co-Au and Co-Co obtained from the best EXAFS simulations (see figure 2, right), evidences a sharp Co/Au interface with no interdiffusion in such immiscible couple.

Getting back to Table 1, despite a significant progressive increase in the ZFC peak temperature ($T_{max}$) and in the coercive field ($\mu_o H_c$) from Co@C, Co@Cu and Co@Au, we found that the effective MAE distribution, sensitive to the interface hybridization, does not increase in the same proportion. Moreover, there is no modification (within the error bars) of the constant ratio ($K_2/K_1$) close to 1.2 whatever the matrix nature. One can conclude that the magnetic anisotropy of such nanoparticle assemblies seems to be dominated by the shape and crystal structure of the particle surface (i.e. additional or incomplete facets) rather than by volume or environment as described in (Jamet et al. 2001), (Oyarzun et al. 2015) and (Xie et Blackman 2004).

Then, by increasing the volume concentrations in such cluster assemblies, one can modulate the mean distance between NPs and thus investigate remaining open questions concerning the long range effects of interactions between nanomagnets versus matrix nature (Hillion et al. 2017). Indeed, for an assembly of randomly oriented independent macrospins with 1% volume concentration, the edge-to-edge mean clusters distance is greater than 7nm and no magnetic interaction are expected in agreement with $\Delta m$ found around zero. But with 3% volume concentration, the edge-to-edge mean clusters distance is as little as 2nm and one can then expect significant interactions between neighboring NPs. As an example, on Co@Au samples with 3% volume concentration, SQUID magnetometry measurements samples revealed an increase of the peak temperature ($T_{max}$) of ZFC curves up to 30 K and a negative peak $\Delta m$ up to 10%, related to demagnetizaing interactions. By increasing concentrations up to 4%, we observed that $T_{max}$ and $\Delta m$ continue to increase, while the coercive field ($\mu_o H_c$) remain almost unchanged compared to the previous 3%-concentration. Moreover, as the magnetization at high temperature is essentially sensitive to the magnetic size distribution, $m(H)$ curves above $(T_B)$ have been found clearly concentration dependent. From the best fit to reproduce the ZFC/FC and $m(H)$ at 300K, we have estimated that the magnetic dimers proportion represent 15% (resp. 20%) in 3% (resp. 4%)-Co@Au sample. Thus, we have proposed a simple model where *magnetic dimers* are formed for distances lower than a given magnetic interaction length *l\**. This distance *l\** has been found equal to 1.2 ± 0.2 nm in Au matrix, much smaller for non-metallic Ge and C matrix but similar for Co@Cu samples, probably because Ruderman-Kittel-Kasuya-Yosida (RKKY) Interactions are the major ingredient of *superferromagnetic* dimerization through the conduction electrons of metallic matrix. Note that such RKKY interactions are previously evidenced from giant magnetoresistance (GMR) in granular Co@Cu samples (Oyarzun et al. 2013).

As a conclusion, this approach where the very same Co nanoparticles are diluted in different matrices with independent control of size and concentration, provides an original and quite unique way to experimentally probe interface magnetic anisotropies but also interparticle interactions between nanomagnets from phenomenological methods. Nevertheless, for higher concentrations, long-range dipolar interactions may play a major role and collective effects can occur so that much more complex models are needed.

*Assemblies of Co-clusters in MgO matrix*

On the contrary, when Co clusters are embedded in MgO matrix even for low dilution, atypical magnetic behaviour is revealed with high ZFC peak temperature ($T_{max}$=71 K) and high coercive field ($\mu_o H_c$=400 mT) (see figure 3). Indeed from a "triple fit", $K_1$ reaches 620 kJ.m$^{-3}$ for 1% Co@MgO as-prepared sample and is some 4-5 times higher than the MAE obtained for previous matrices. And in this case the biaxial fit does not apply for magnetization curve at low temperature (see Fig. 3.b).

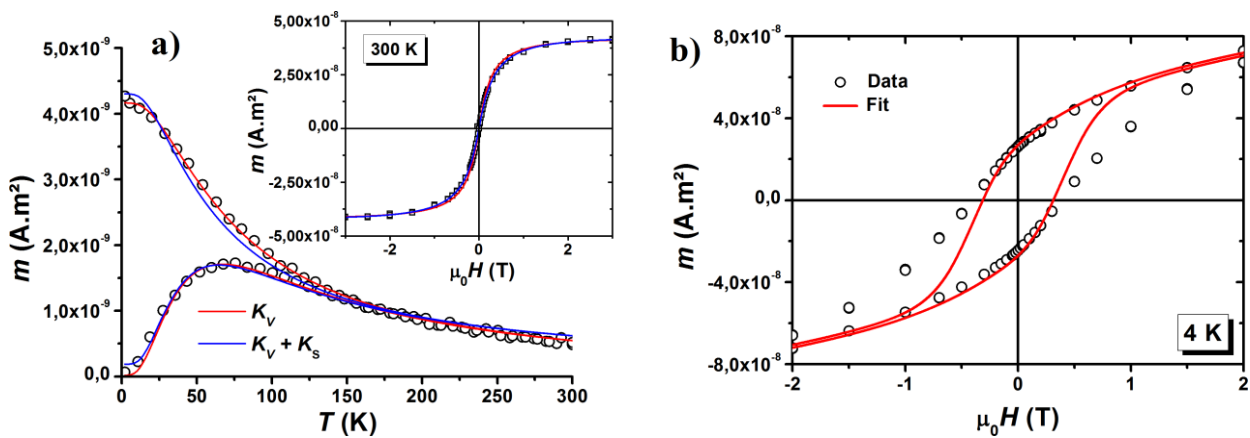

**Fig. 3:** ZFC/FC susceptibility and magnetization curves at 300 K (in insert) (a) and hysteresis loops at 4 K (b) for Co nanoparticles embedded in MgO matrix. The red solid line is a tentative fit using the same set of parameters with biaxial fit as for previous samples while the blue solid is the fit by taking into account volume and surface contributions.

To go further, we analyse SQUID measurements by taking into account non-negligible interface effects and write the cluster MAE as the summation of volume and surface terms as $\Delta E = K_v V + K_s S$.

In order to give a quantitative evaluation of the matrix influence on the MAE, the ZFC response of a clusters assembly with a random orientation of the anisotropy axis, can be further described as a function of temperature by (Anderson at al. 1997):

$$M_{ZFC}(T) = \frac{\mu_0 \mu^2 H}{3K}\left[1 + \frac{K}{k_B T}\left(1 - e^{-t/\tau(T)}\right)\right]$$

where $\mu$ is the cluster magnetic moment and $t$ is the time measurement (about 60 s in our case). Using the convolution of this equation with the magnetic size distribution, the fits for Co@MgO sample (presented in blue in Fig. 3) allow us the determination of $K_V$ and $K_S$ compared to ones for Co@Nb sample, as reported in Table 1.

| Sample | $T_{max}$ | $D_{mag}$ | $K_V$ | $K_{S,i}$ | $K_{S,inter}$ |
|---|---|---|---|---|---|
| Co@Nb | 12 | 2.27 | 60 | 45 | 0 |
| Co@MgO | 100 | 2.30 | 30 | 45 | 265 |

Table 1: Blocking Temperature ($T_{max}$ (K)), mean magnetic Diameter ($D_{mag}$(nm)), Volume-anisotropy energy ($K_V$ (kJ/m$^3$)), Surface ($K_{S,i}$ ($\mu$J/m$^2$)) and Interface ($K_{S,\,inter}$ ($\mu$J/m$^2$)) anisotropy energy for Co@Nb and Co@MgO sample, determined from the ZFC curves simulations (Rohart et al. 2006).

To discuss the surface anisotropy variation, $K_S$ can be seen as the summation of both intrinsic surface cluster $K_{S,i}$ and the specific cluster/matrix interface $K_{S,inter}$ contributions. As a rough approximation, we assume that the interface anisotropy is negligible in Nb matrix (Rohart et al. 2006). So in Co@Nb samples, the surface anisotropy is only limited to the intrinsic cluster surface contribution as evidenced by micro-SQUID measurement on single NP (Jamet et al. 2001). The important $K_{S,inter}$ observed for the clusters embedded in the MgO matrix (see table 1), originates from a partial oxidation of the cluster surface. This leads to the formation of an antiferromagnetic (AFM) shell around the cluster which causes an exchange anisotropy (Dobrynin et al. 2005) as in Co/Co$_{1-x}$Mg$_x$O bilayers (Hong et al. 2006).

Even if such huge MAE has been proposed to be used to beat the *superparamagnetic* limit by exchange bias between a metallic Co core and CoO anti-ferromagnetic shell (Skumryev et al. 2003) (Le Roy et al. 2011), it may be noted that after a few months by ageing in air, oxidation probably extend with severe evolution of the magnetic signal (see figure 4). In particular, the coercive field decrease and an exchange bias of 20 mT have been observed one year later by measuring Field-Cooled m(H) curves at low temperature (see Fig. 4b).

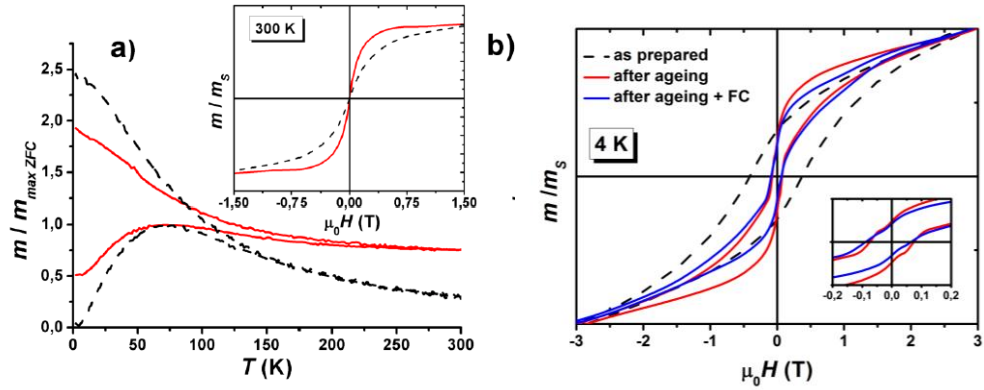

**Fig. 4:** Comparison of ZFC/FC susceptibility and magnetization curves at 300 K (in insert) (a) and hysteresis loops at 4 K for as-prepared (black dashed) and aged sample (with ZFC in red and FC in blue m(H) solid curve) Co nanoparticles embedded in MgO matrix (b).

In order to distinguish intrinsic and extrinsic parameters influence on the cobalt magnetic moment, X-ray circular magnetic dichroism (XMCD) measurements at the $L_{2,3}$ cobalt edges have been performed less than one month later on such Co@MgO sample. One can mention that the $L_{2,3}$ cobalt edge is essentially metallic with a shoulder corresponding to small oxide part and that the hysteretic loop measured at 300 K is very closed to the one measured by SQUID after ageing (see Fig.4 and 5). By applying the nowadays well-known sum rules (Carra et al. 1993, Thole et al. 1992), we determine a spin magnetic moment equal to 1.28 $\mu_B$ per atom to compare to 1.62 $\mu_B$ per atom for the Co-bulk phase while the orbital contribution is very low 0.056 $\mu_B$/at to compare to the 0.15 $\mu_B$ per atom for the Co-bulk phase. We can claim that this moment reduction of 20 % comes from a partial cluster oxidation which induces lower than one non-magnetic monolayer in the case of Co clusters embedded in MgO.

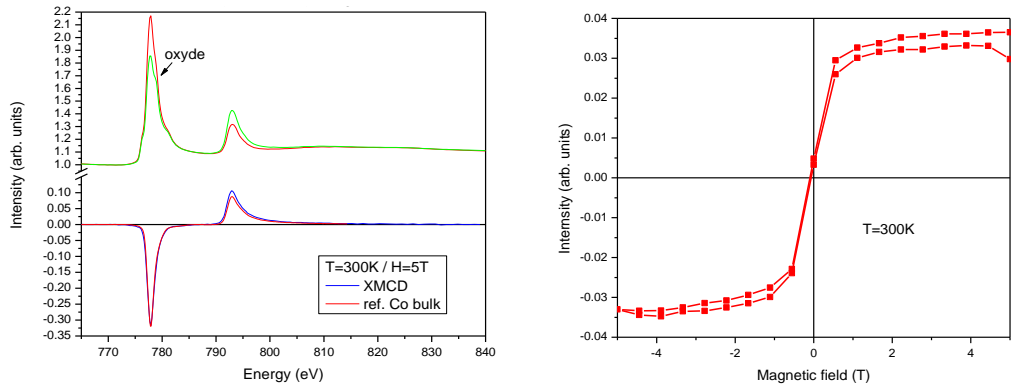

**Fig. 5**: XMCD measurements on the Co/MgO sample at the cobalt $L_{2,3}$ threshold (a) and magnetization curve measured at 300 K following the normalized Co $L_3$ edge absorption intensity as a function of applied magnetic field (b).

It must be noted that XMCD measurements also revealed that Co atoms are partially oxidized for CoPt alloy clusters embedded in MgO matrix, with a $\mu_s$ decrease but a $\mu_L/\mu_s$ enhancement (Tournus et al. 2008). Thus, the MgO environment appears to have a strong effect on the magnetic properties also for bimetallic particles such as CoPt or CoAu (see Fig. 6). The cluster/oxide interface can in this case induce a different chemical arrangement in the particles, compared to other matrices, in addition to a decrease of the particle magnetic size (dead layer) and an enhancement of the magnetic anisotropy. The ZFC susceptibility measurements shown in Fig. 6, for the same particles (CoPt or CoAu) diluted in

C and MgO illustrate this strong interface effect of MgO, which results in a significant enhancement of the peak temperature ($T_{max}$, which is related to the blocking temperature).

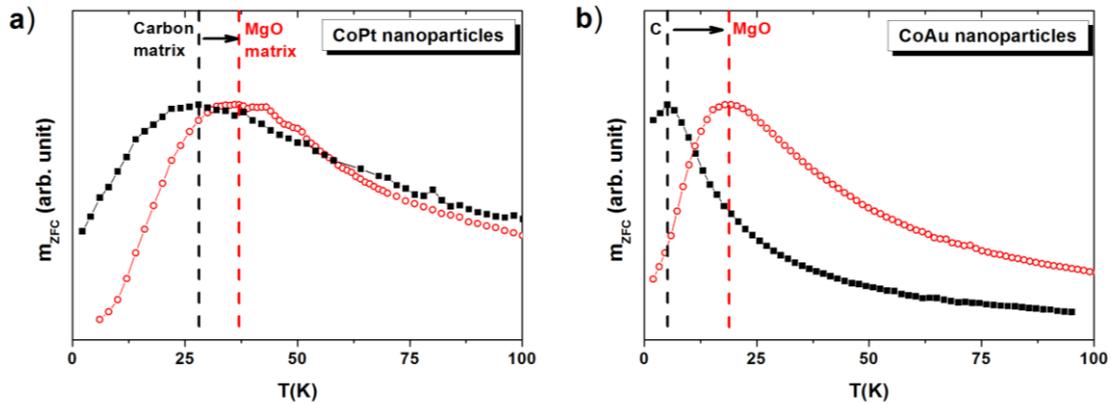

Fig. 6: ZFC susceptibility measurements for CoPt nanoparticles (a) and CoAu nanoparticles (b) diluted either in an amorphous carbon matrix or a MgO matrix. The particles appear to be blocked up to a higher temperature in the case of a MgO matrix.

## II- Order/Segregation in 3d-5d TM nanoparticles

Over the last decade, bimetallic magnetic nanoparticles (NPs) have attracted considerable attention as potential candidates for various applications from catalysis, magnetism, optics, to nanomedicine (Calvo et al. 2013). From a fundamental point of view, due to intricate size and hybridization effects between two different TM-species, the magnetic properties of nanoalloys have been intensively studied both theoretically and experimentally by varying the composition, degree of chemical ordering and morphology. We will focus our attention on 3d-5d TM bi-metallic NPs with two opposite tendencies in bulk-phase, namely to form alloyed phase (as in CoPt or FePt) or phase-separation (as in CoAu or FeAu). At nanosize, as the phase stability becomes more complex, we can mention that reversely to equilibrium thermodynamics predictions, high fraction of surfaces in NPs can promote partial segregation as in FePt NPs (Pohl et al. 2017) or ordering tendency as $L1_0$ superstructure in heat-treated FeAu NPs (Mukherjee et al. 2012). In these part, we report on experimental studies of such bimetallic NPs prepared by LECBD and embedded in amorphous carbon matrix.

*Assemblies of Pt-based magnetic clusters in matrix*

First of all, alloys made of iron or cobalt (Fe or Co) and platinum (Pt) are always of particular interest because they can yield to one of the magnetically 'hardest' materials (magnetization is retained for a long time) as necessary for high density media storage applications (Andreazza et al. 2015). As their equilibrium phase diagram in bulk-alloy is very rich, mass-selected equiatomic CoPt nanoparticles (Blanc et al. 2013) and FePt nanoparticles (Dupuis et al. 2015) has been produced by laser vaporization cluster source and annealed under vacuum in order to promote chemical ordering. The atomic structure of CoPt and FePt NPs with a diameter between 2 and 5 nm, has been studied by advanced transmission electron microscopy (Tournus et al. 2013). In addition to particles corresponding to a single $L1_0$ bulk-like ordered domain, we put into evidence that even small particles can display multiply twinned particles with decahedral or icosahedral shapes but that the chemical order can be preserved across twin boundaries. By combining high photon flux and chemical selectivity of SR facilities, XMCD have been used to study magnetic moment of CoPt and FePt clusters assemblies before and after transition to the chemically $L1_0$–like phase. As the enhanced proportion of low coordinated atoms at the surface (which corresponds to around 40 % in the 3nm size–range) causes a narrowing of the valence d band inversely proportional to the density of state at the Fermi level, we found significant increase compared to the bulk, both for the spin and orbital moments of Fe, Co and Pt atoms in such nanoalloys. We have also studied size effect on the local structure and MAE in diluted assemblies of CoPt clusters embedded in a carbon matrix upon annealing in the 2–4 nm diameter range. From Co-K and Pt-L edges EXAFS experiments and simulations, we evidenced an element-specific dependence of the local atomic relaxations in CoPt clusters leading to a strong distortion in pure Co planes since pure Co layers do not match the underlying Pt layer in chemically ordered $L1_0$-like clusters, in agreement with ab-initio VASP simulation calculations (Blanc et al. 2013). We claimed that such structural distortions added to the statistical chemical distribution and exotic structures (with five-fold symmetry) in annealed CoPt NPs, could explain the low effective MAE normal distribution increase of only 35 % compared to the one of as-prepared sample measured by advanced SQUID magnetometry (Dupuis et al. 2013). This value has been found one order of magnitude smaller to what is expected for the $L1_0$ bulk CoPt phase. Only recently, (Yang et al. 2017) quantitatively demonstrate correlated local defects (in chemical order, composition…) and surface-relaxation effects to magnetic anisotropy energy (MAE) decrease by a state-of-the-art approach at the single-atom level combining atomic electron tomography correlated to advanced DFT calculations on FePt nanoparticles embedded in carbon.

*Assemblies of Au-based magnetic clusters in matrix*

Multifunctional applications exploiting magneto-plasmonic properties are expected by combining 3d TM (Fe or Co) and noble metal (Ag and Au) (Maksymov 2016). But, in the CoAu and FeAu case, due to limited solubility and positive heat of mixing (HOM), phase separation for a wide range of composition is reflected in equilibrium bulk-phase diagram. Thus such 3d-5d TM NPs could present core-shell morphologies. Notice that iron NPs passivated by gold coating favored for its specific surface functionality to fix chemical or bio-medical agents, are expected to play an important role in the wide range of sophisticated bio-medical applications such as targeted drug delivery, biochemical sensing, and ultra-sensitive disease detection (Ban et al. 2005). Nevertheless at nanosize, (Mukherjee et al. 2014) claimed that the stabilization of compounds or phases, arise from a competition between both size-dependent HOM and total surface energy (SE) of the NP and the interphase interfaces created. So by decreasing the size, as the HOM decreases more rapidly than SE, the miscibility can increase. They argue that the $L1_0$ ordered FeAu is stable below 10 nm even if the stability factor *($\Delta F = SE - HOM)$* is near-zero at such nanosize.

So we have synthesized Au-based NPs from LECBD in the size range 2 to 5 nm in diameter to study the structure and morphology from high resolution transmission electron microscopy (HRTEM) experiments (Dupuis et al. 2015 b). On CoAu NPs coated by amorphous carbon, we experimentally evidenced a thermal transition, as illustrated in figure 7 with HAADF images and associated STEM-EELS core-loss chemical mapping of CoAu NPs before and after annealing (recent observations performed on the NION USTEM 200 at the LPS, Orsay in collaboration with K. March and M. Kociak). The final core/shell structure with off-center Co core is in remarkable agreement with theoretical calculations (Palomares-Baez et al. 2017). The structure is segregated in both case (respectively (Co shell) for as-prepared and (off-center Co core) for annealed CoAu NPs, no alloyed phase has ever been observed.

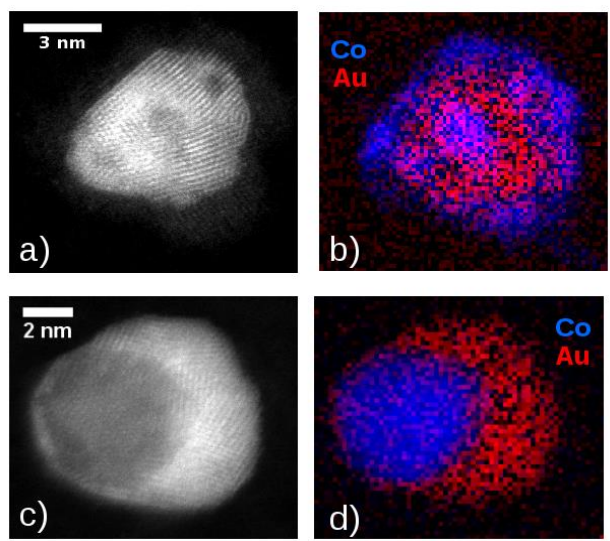

**Fig. 7:** HAADF images of CoAu Nps with associated EELS core-loss chemical mapping (on gold M and cobalt L ionization edges): a) and b) as prepared CoAu nanoparticle, c) and d) annealed CoAu.

On the contrary on as-prepared FeAu NPs, fcc-phases have been observed from HRTEM with lattice parameters varying between 3.7 to 4.08 Å (see Fig.8a and 8b). While, a Fe core/Au Shell has been revealed from Z-contrast imaging, generated by high angle annular dark field (HAADF) scanning transmission electron microscopy (STEM), as the first clear evidence of possible segregation in a so small FeAu NP seen in (Fig. 8c). But upon annealing, contrary to CoAu NPs, we systematically obtained fcc alloyed phases for FeAu NPs and anomalous chemically $L1_0$ ordered on one heat-treated FeAu NP as Mukherjee et al. 2012 (see Fig. 9). In their theoretical paper, (Zhuravlev et al. 2017) predicted that $L1_0$ FeAu, $L1_2$ FeAu$_3$ and $L1_2$ FeAu$_3$ are unstable in density functional theory (DFT) calculations. They assume that an indirect influence of compressing effects at the surface which lead to concentration waves at the corresponding [001] ordering, with enrichment of Fe in the core and segregation of Au to the surface, could explain the ordering tendencies in 5nm-FeAu nanoparticles. They also evaluate the possible role of different type of magnetic order in FeAu clusters, where the specific different patent lattices (as bcc in Fe and fcc in Au) compete with each other and the ordering of alloys containing such 5d metals may also be affected by the Spin-orbit coupling (Zhuravlev et al. 2017).

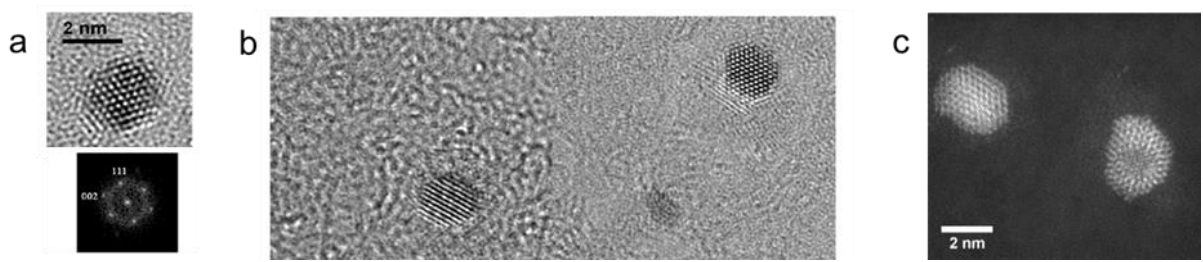

**Fig. 8:** HRTEM observations on as prepared FeAu NPs showing single fcc phase with a=3.7 Å a) and with a=4.08 Å b) as lattice parameter and STEM-HAADF observations with fcc and Fe Core/ Au Shell c) cohabitation.

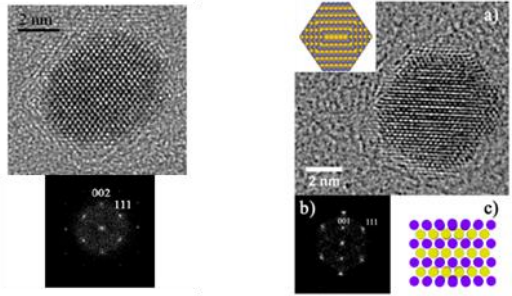

**Fig. 9:** HRTEM observations on annealed FeAu NPs with single fcc phase (left) and L1$_0$ FeAu (right) nanoalloys with 3D representation of the chemical ordered Wulff truncated octahedron in inset (a), the typical (001) sur-structure peak in the fast FT of the HRTEM (b) and the corresponding chemical order stacking (c)

To go a step further experimentally, we prepared assemblies of FeAu embedded in carbon to statistically study their structure from EXAFS measurement at both edges (K-Fe and L-Au as-previously for pure clusters at the CRG-BM30b-FAME beam line of the ESRF) and their magnetic behavior (both from Squid magnetometry and XMCD at L-Fe edge on X-Treme of the SLS and at L-Au edge on the ID12 beam line of the ESRF) before and after annealing.

At the Fe-K edge, the FT EXAFS obtained for FeAu@C sample before and after annealing signal is completely different from the bcc-bulk Fe phase but very similar to the signal obtained on previous fcc-Co NPs embedded in Au and C matrix in Fig. 2. Upon annealing, we observe no significant differences from the shape and amplitude of the FT EXAFS signature at the K-Fe edge. On the contrary at the L-Au edge, both the NN amplitude and distance increase while the Au-C bonding and the Debye-Weller disorder decrease after thermal treatment. Moreover, the average lattice parameter in annealed FeAu samples tends to the Vegard law value of 3.85 Å (between $a_{fcc\text{-}Au}$=4.07 Å and $a_{fcc\text{-}Fe}$= 3.66 Å), in agreement with 59% at. (±5%) Fe concentration found from Rutherford Back Scattering (RBS) measurements (see Fig. 10). From SQUID magnetometry measurements (see Figure 11) and simulations, we observe no significant thermal evolution of the global magnetic behaviour where the magnetic diameter is the TEM one with constant $T_{max}$ and $\mu_o H_c$ respectively equal to 12 K and 45 mT with moderate anisotropy kept constant around 100 kJ/m$^3$. From XMCD measurements at both edges, the high-Spin fcc iron state and the induced magnetic moment in gold slightly evolve after annealing up to a total magnetic moment of 2.12 $\mu_B$ per Fe atom and 0.09 $\mu_B$ per Au atom, respectively (see Fig. 12 and Table II). Our accurate experimental results are in very good agreement with first-principle calculations where the ferromagnetic order is energetically to be the most favorable configuration with induced magnetic moment for Au atoms of the order of 0.05 (Mukherjee et al. 2012) as measured at the Au-L edge (Dupuis et al. 2015 b). Notice that no signature of antiferromagnetic ordering has been observed on our annealed FeAu@C samples, reversely to the very low global magnetization and low coercivity values measured on FeAu NPs with a smaller lattice parameter of the order of 3.7 Å by (Mukherjee et al. 2012).

As a conclusion, due to finite size effects, we have demonstrated that as-prepared nanoparticles by LECBD can sustain metastable compound structure at the interface between pure cluster and matrix (as in Co@C sample) but also inside alloyed clusters (as in FeAu nanoparticles). We have shown that heating treatments allow to reach stable solid phase up to non-equilibrium chemically ordered phase in nanoalloy, probably to relate to specific atomic relaxation (Blanc et al. 2013). Such distortions can strongly modify the MAE (as in L1$_0$-CoPt) but energetically promote FM order at nanosize leading to an increase of the intrinsic magnetic moment in supported 3d TM clusters and to an induced spin moment when alloyed with 5d but also with 4d TM as previously shown in B2-FeRh Nanoparticles (Hillion et al. 2013 b).

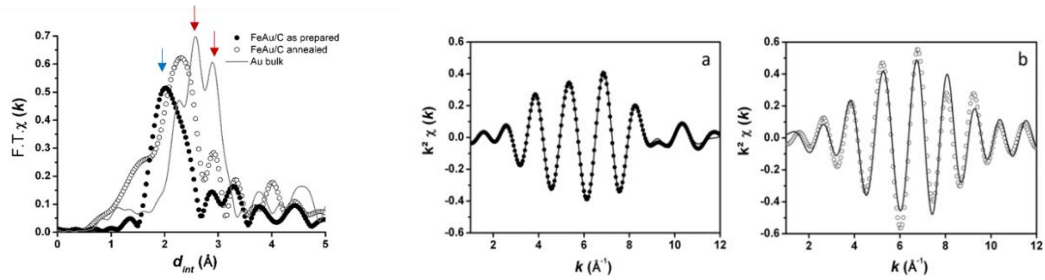

**Fig. 10:** Fourier Transform of the EXAFS signal for FeAu clusters embedded in C (as-prepared and annealed) at the L-Au edge (left). Comparison between the experimental EXAFS signal (dots, contribution of the nearest neighbours (NN) peak only) and simulated curves (solid lines) at the L-Au edge for FeAu sample before and after annealing (right).

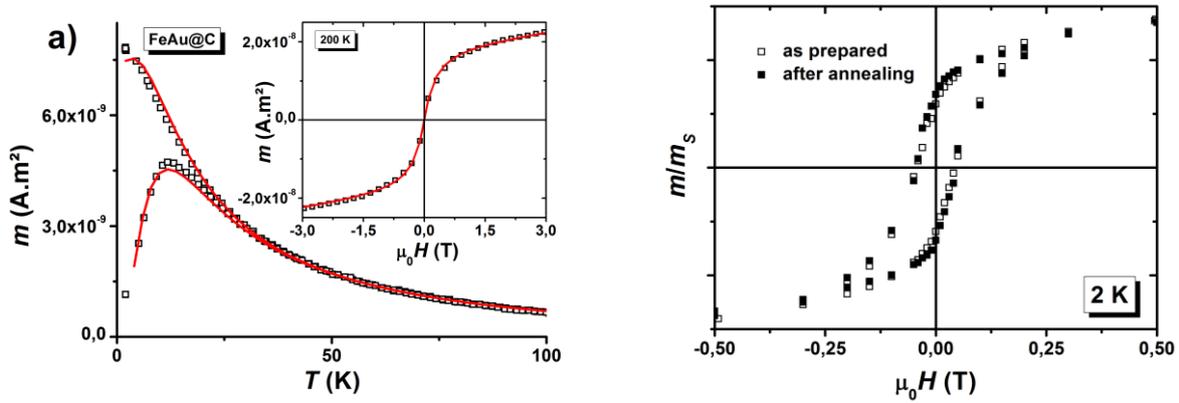

**Fig. 11:** ZFC/FC susceptibility and magnetization curves at 200 K (in insert) (a) and hysteresis loops at 4 K (b) for FeAu nanoparticles embedded in C matrix. The red solid line is the best triple fit adjustment.

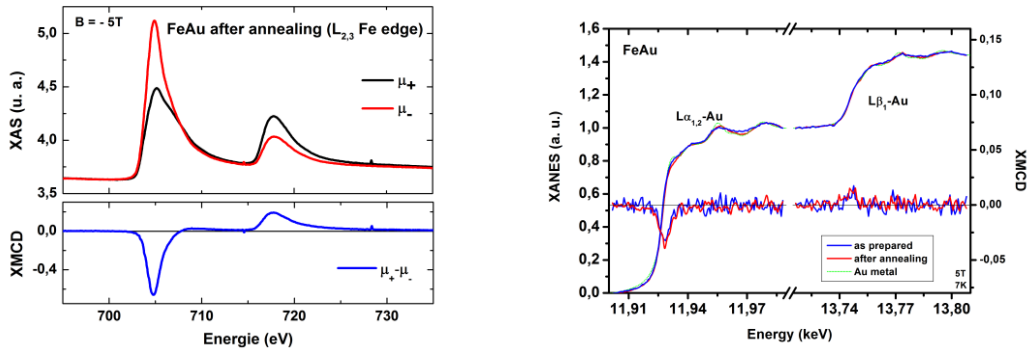

**Fig. 12:** Comparison between the XMCD signal at the Fe-L edge and at the Au-L edge obtained on annealed mass-selected FeAu clusters embedded in carbon matrix with 3 nm in diameter.

| Fe-L edge | $m_S$ ($\mu_B$/at.) | $m_L$ ($\mu_B$/at.) | $m_L/m_S$ |
|---|---|---|---|
| FeAu as-prepared | 2.07±0.3 | 0.02±0.02 | 0.01±0.02 |
| FeAu annealed | 2.04±0.3 | 0.08±0.05 | 0.04±0.03 |
| **Au-L edge** | $m_S$ ($\mu_B$/at.) | $m_L$ ($\mu_B$/at.) | $m_L/m_S$ |
| FeAu as-prepared | 0.04±0.02 | 0.02±0.02 | 0.4±0.2 |
| FeAu annealed | 0.07±0.03 | 0.02±0.02 | 0.24±0.1 |

**Table II**: Spin and orbital moment of Fe and Au from XMCD measurements at both edges on FeAu clusters assemblies embedded in C matrix with 2.9 nm in diameter.

### III- Prospective for the next years.

Today, the state-of-the-art progress at the single-atom level combining for example atomic electron tomography correlated to advanced DFT calculations, allow to accurately describe the intrinsic properties of more and more complex magnetic clusters assemblies (Yang et al. 2017). *In operando* experiments are also allowing to describe extrinsic properties of clusters assemblies near real-life conditions or in reactive atmosphere taking into account their physico-chemical surface affinities (as Ramade et al. 2017) which is in particular interest to optimize their use as catalyzers. Looking forward, the ability to determine crystal defects with high precision in nanoalloys and to functionalize hybrid magnets at the nanoscale opens up the possibility for revolutionary new applications in many fields of science and industry.

Nowadays, there are high hopes for nanomedicine applications by exploiting the ability to functionalize magnetic NPs with biomolecules, as well as to manipulate them by applying external magnetic fields (Sandhu et al. 2010). Progresses for the health are expected from theranostic magnetic biocompatible nanoparticles (Gobbo et al. 2015). From medical imaging, drug nanovector delivery and tumor destruction by hyperthermia to magnetic cell separation, magnetofection, and even magnetic control of cell signaling pathways, magnetic micro- and nanoparticles are finding ever increasing use in biomedicine and biomedical research (Gleich and J. Weizenecker 2005), (Gijs et al. 2010) and (Connord et al. 2015).

There is also a growing demand for high performance magnets, notably for green energy such as electric motors and generators. Until now, the breakthroughs in magnet research have been based on NdFeB but the monopoly of such compound causes a problem of heavy rare earth (RE) resources. One solution could be the use of RE nanoscopic objects (Schmidt et al. 2015). Another promising route could be to fabricate nanocomposite materials made of fine mixture of a hard and a high magnetization phase in strong magnetic coupling. Theoretical models on hard-soft permanent-magnet nanocomposites, predict superior performances than the current permanent magnets (Skomski et al. 2014).

Technological improvements for ultra-high density magnetic storage applications always still require the understanding of magnetic anisotropy energy at the nanometer range and of dynamical magnetization reversal processes at the nanosecond time scales of benchmark nanomagnets.

In a next future, we propose to explore alternative magnetic nanoparticles prepared by MS-LECBD in interaction with their environment, to study size effects, strain on monocrystalline template or in matrix confinement to tune their extrinsic magnetic properties.

As an example (Linas et al. 2015) has pointed out the possibility to self-organize Pt nanoclusters by MS-LECBD on a moiré pattern from the epitaxy of graphene on Ir(111), similar studies are in progress with network of hard $L1_0$ FePt NPs to study charge transfer effects and electronic structure modifications on their magnetic properties.

Controlling nanomagnetism by means of an electric field is a key issue for the future development of low-power spintronic (Chappert et al. 2007). Recently, a ferroelectric $BaTiO_3$ crystal has been used to electrically drive the metamagnetic transition temperature of epitaxially grown FeRh films with only a few volts, with potential media applications for Thermally Assisted magnetic Recording (Cherifi et al 2014). By increasing the materials confinement at the ultimate nanoscale, we propose to try to modulate in real time, (AFM-FM) order transition in CsCl B2-FeRh nanoparticles with adjustable nanosize, using an electric field by combining effects of screening charges accumulation and epitaxial strains at the interface with a ferroelectric and piezoelectric $BaTiO_3$ crystal. Once demonstrated, the same principle and device could be used to modify interface magnetic exchange coupling existing between a FM nanoparticle and a surrounding dielectric, AF material, like CoO and heavy Pt metal via spin–orbit torque as in magnetic skyrmions of new spin configuration which are anticipated significantly more energetically stable (per unit volume) than their single-domain counterparts (Schott et al. 2017) (Romming et al. 2015). With the ultimate goal of characterizing single nanomagnets it will be a first step towards the development of nanodevice for electronic transport through a nanocontact (Misiorny et al. 2015).

**Acknowledgements**

All the clusters samples were prepared in the PLYRA [a] platform, created in the frame of project agreement 1989-1993 between the french State and the Rhône-Alps Region, as a joint initiative of A. Perez, M. Broyer and A. Renouprez from 3 laboratories of Lyon (LPMCN : Laboratoire de Physique de la Matière Condensée et Nanostructures – UMR UCBL-SPM CNRS N° 5586 - LASIM : Laboratoire de Spectrométrie Ionique et Moléculaire – UMR UCBL-SPM CNRS N° 5579 - IRCELYON : Institut de Recherche sur la Catalyse – UPR Chimie CNRS N° 5401). Many thanks to them and to the technical staff G. Guiraud, F. Valladier and C. Clavier, startup responsibles of the building and development of the laser vaporization cluster machines at the Université Claude Bernard Lyon 1 (UCBL) but also to all the PhD-students and post-doc researchers who have participated to the adventure. SQUID measurements were first performed at the Néel institute then at the CML [b] platform created in 2008 at UCBL. The authors are also grateful to A. Ramos, H. Tolentino, M. de Santis and O. Proux for their help during XRD and EXAFS experiments on the French CRG-D2AM and BM30b-FAME beamlines at ESRF, to S. Rusponi and H. Brune from EPFL for stimulating discussions, J. Dreiser, C. Piamonteze and F. Nolting from Swiss Light Source for their investment on the X-Treme beamline, P. Ohresser from DEIMOS beam line SOLEIl synchrotron and K. Fauth from University of Wuerzburg for their help at BESSY II experiments, A. Rogalev and F. Wilhelm from the ESRF ID12 beamline for the XMCD measurements. Support is acknowledged from the French national CNRS ACI, METSA network, GDR and ANR on clusters and Nanoalloys and from the European Community AMMARE contract No G5RD-CT 2001-0047P, STREP SFINx No. NMP2-CT-2003-505587 and COST-STSM-MP0903 on Nanoalloys.

[a] Plateform Lyonnaise de Rechecherche sur les Agrégats, Université Lyon 1-CNRS, Université de Lyon, 69622 Villeurbanne cedex, France
[b] Centre de Magnétomètrie de Lyon, Université Lyon 1-CNRS, Université de Lyon, 69622 Villeurbanne cedex, France

**Conflict of Interest**: The authors declare that they have no conflict of interest.